\documentclass[twocolumn,prX,showpacs,showkeys]{revtex4}
\usepackage{latexsym}
\usepackage{graphicx}
\newcommand{\x}{\vec x}
\newcommand{\y}{\vec y}
\newcommand{\xt}{(\x,t)}
\newcommand{\phiop}{\hat\varphi}
\newcommand{\avr}[1]{\left\langle{#1}\right\rangle}
\newcommand{\intk}{\int\frac{d^dk}{(2\pi)^d}}
\newcommand{\ak}{{\hat a_{\vec k}}}
\newcommand{\kx}{{\vec k\vec x}}
\newcommand{\psik}{{\psi_{\vec k}}}

\renewcommand{\L}{{\cal L}}
\begin{document}
\title{Semiclassical decay of topological defects}
\author{Szabolcs Borsanyi}
\email{s.borsanyi@sussex.ac.uk}
\author{Mark Hindmarsh}
\email{m.b.hindmarsh@sussex.ac.uk}
\affiliation{Department of Physics \& Astronomy, University of Sussex,
Brighton BN1 9QH, UK}


\begin{abstract} 
Perturbative estimates suggest that extended topological defects such 
as cosmic strings emit few particles, but numerical simulations of the 
fields from which they are constructed suggest the opposite. 
In this paper we study the decay of the two-dimensional prototype
of strings, domain walls in a simple scalar theory, solving
the underlying quantum field theory in the Hartree approximation. 
We conclude that including the quantum effects makes the picture clear:
 the defects do not directly transform into particles, but there
is a non-perturbative channel to microscopic classical structures in the form 
of propagating waves and persistent localised oscillations, which 
operates over a huge separation of scales.
When quantum effects are included, the microscopic 
classical structures can decay into particles.
\end{abstract} 

\maketitle

\section{Introduction} 
Our current fundamental theory of the universe suggests higher order
symmetries which are successively broken as the universe cools. This naturally
leads to the formation of topological defect networks \cite{Kibble:1976sj}. Such
extended objects could play a role in the early formation of structure in the
universe \cite{Vilenkin:1983jv,KibbleHindmarsh,VilenkinShellard}. 
Indeed, recent calculations \cite{Bevis+al}
show that models with cosmic strings and other topological 
defects fit the Cosmic Microwave Background data better than the standard 
power-law $\Lambda$CDM model.


Analytical estimates suggest that particle production from decaying strings
(or, in two dimensions: domain walls) is suppressed by the separation of
cosmological ($\ell$) and microscopic ($M$) scales. The former is the curvature
scale of the string, assumed to be of order the Hubble radius, and the
latter is determined by the underlying particle physics, typically on the GUT
scale.  Particle production would mean transforming the energy in the deep
infrared into ultraviolet excitations, over something like 58 orders of magnitude in 
momentum scale.  If the cosmic string network is
prohibited from losing energy, its dynamics follows the Nambu-Goto action
\cite{NambuGoto,KibbleHindmarsh}. Perturbation theory suggest that the radiative
energy loss is $\sim M/\ell$ \cite{Srednicki:1986xg}. With more insight into the
non-linear domain wall dynamics one discovers another channel 
through cusp annihilation \cite{Brandenberger:1986vj}, at a rate of $\sim M^{5/3}\ell^{-1/3}$. 
Once one takes into account the gravitational channel a simple estimate gives $\sim
10^2M^4/M_p^2$, where $M_p$ is the Plack mass \cite{Gravidecay}, which 
therefore seems to be dominant for sufficiently large $\ell$.


Numerical analysis of the
Nambu-Goto action confirms \cite{AlbrechtTurok,Bennett:1987vf,Allen:1990tv} the analytical scaling
assumption \cite{Kibble:1984hp} that implies a string density $M/t^{d-1}$
(in $d=2,3$ space dimensions), at least for strings as long as the horizon size. 
An exact Minkowski space simulation of a string network
reveals that fragmentation to loops is the dominant decay process \cite{Smith:1987pv,Sakellariadou:1990nd},
which proceeds down to the smallest scale on the network, which may 
even be the microscopic scale of the string width
\cite{Vincent:1996rb}.  Simulations in an expanding universe \cite{Bennett:1987vf,Allen:1990tv} 
broadly confirm this 
picture, although indicate that the relevant small scale is the initial correlation length 
\cite{Ringeval:2005kr,Martins:2005es,Olum:2006ix}\footnote{It should 
be noted that there is apparent disagreement between the three sets of authors on the interpretation of 
their results.  We base our assertion on the observation that the peaks of the loop production function 
and the loop distribution function remain near the initial correlation length.}


However, for strings which are topological defects, we can check this picture 
by solving the underlying field theory in the classical approximation. This
means in practice to integrate the non-linear wave equations on a spatial
lattice and to average over an initial ensemble. This approach offers a full
insight into all non-perturbative phenomena, although it may be difficult to
justify the omission of quantum effects at the microscopic scale.
Nevertheless, this method has been successfully used elsewhere, e.g. to explore
the dynamics of symmetry breaking \cite{Felder:2000hj}, or of non-thermal phase
transitions \cite{Kofman:1995fi} in the post-inflationary Universe.

The scaling behaviour suggested by Nambu-Goto simulations is manifest
in the classical field dynamics, even though only the microscopic scale
appears in the equation of motion. It has been demonstrated in the context
of gauge strings in the Abelian Higgs model \cite{Vincent:1997cx,Moore:2001px}
global strings \cite{Yamaguchi:1999,Moore:2001px}, 
non-Abelian global strings with junctions \cite{Hindmarsh:2006qn},
and 
semilocal strings \cite{Achucarro:2007sp,Urrestilla:2007sf},
as well as domain walls \cite{Garagounis:2002kt,Oliveira:2004he}, 
including models with junctions \cite{Avelino:2006xf}.
The scaling is present in Minkowski as well as expanding space time, and in
two as well as three dimensions.  It seems to be a universal feature of 
classical field theories with extended structures.

For strings, a major difference between the numerical solutions of the
classical field dynamics and Nambu-Goto simulations is that defects decay into
classical radiation \cite{Vincent:1997cx}, at a much faster rate than
anticipated from perturbation theory and cusp annihilation.  One typically
observes a length of string $\ell$ in a volume $\ell^d$, and hence that the
string length density is $\L \sim \ell^{1-d}$.  Given a mass per unit length of
$\mu \sim M^2$, the energy density in string is $M^2\ell^{1-d}$.
Since scaling implies $\ell\sim t$, the energy loss rate per unit length
$M^2/\ell$.  Hence, for a loop of size $\ell$, the average energy loss rate is
$M^2$, which is, in fact, greater than the gravitational estimate.

We discover then that the classical scaling implies a strong radiative decay. 
This is very puzzling in view of the scale separation between the $\ell$ and $M$ which 
grows as the simulation proceeds.  However, apart from confirming the scaling 
over more than three orders of magnitude, it is not our purpose to address this 
important question. Instead we note that up to now it has been unclear if 
the classical approach is valid here, where dynamics is driven by an interplay
between macroscopic and microscopic scales. A check for quantum corrections
is crucial.


An alternative to the classical approach is the two-particle-irreducible (2PI)
effective action technique, which is based on a selective resummation of
perturbative diagrams \cite{Berges:2004yj}. Preheating dynamics with
non-perturbative particle production \cite{Berges:2002cz} and particle
thermalisation by scattering \cite{Berges:2001fi} are within the range of its
applicability. The so far used homogeneous version of this elaborate technique
is, however, incapable of addressing the question of defect formation
\cite{Rajantie:2006gy}.

We can combine these techniques, using the classical approach to form defects
and then studying their evolution in the 2PI framework. If we keep the
next-to-leading order diagrams in the 2PI effective action, we will gain
insight into the scattering and thermalisation of the produced particles.  The
inhomogeneous variant of the 2PI equations, however, is technically hardly
feasible. Keeping the lowest order 2PI diagram yields an approximation scheme,
that is equivalent to the well-known Hartree approximation
\cite{Cooper:1987pt,ParisLO}.
While scattering between the produced particles is not included here, even the
homogeneous version of this scheme could account for the non-perturbatively
rapid particle production in the early Universe
\cite{Khlebnikov:1996zt,GarciaBellido:1997wm}. The extension of the equations
to inhomogeneous backgrounds was historically motivated by the hope that the
background field could mediate interaction between the freely streaming
particles. Although numerics have shown that the opposite was true
\cite{Salle:2000hd,AmsterdamInhom,AlamosInhom}, this method can be still used
for finding the leading quantum corrections to the evolution of classical
structures, as has been suggested by a one-dimensional analysis of moving kings
\cite{SalleKinks}.


In this paper we analyse the classical solution of the $\lambda\Phi^4$ theory
in two space dimensions corrected by the Hartree approximation. In the broken
phase this toy model features domain walls, which resemble strings in this
low dimensional setting. We check if there is a significant alteration
to the kink dynamics by the inclusion of this type of quantum correction.

First we recall the results from classical simulations and demonstrate the
scaling behaviour also found in Ref.~\cite{Garagounis:2002kt}.  Then in
section \ref{sec:hartree} we introduce the Hartree approximation of the
considered model. Next, in section \ref{sec:comparison} we numerically
compute the domain wall evolution both in the classical and in the Hartree
approximated framework.  We discuss possible interpretations of the results in
section \ref{sec:discussion}, and finally conclude in section
\ref{sec:conclusion}.

\section{Classical decay of domain walls} 
\label{sec:classical}
\subsection{Model details} 
\label{sec:model}
The Lagrangian density of our scalar theory is as simple as
\begin{equation}
{\cal L}=\frac12 \left[\partial\phi\right]^2-\frac12 m^2\phi^2-\frac\lambda{24}
\phi^4
\end{equation}
The theory has a $Z(2)$ symmetry, this breaks spontaneously if the thermal
mass turns negative.  By the choice of the bare mass parameter $m$ we make sure
that the system is deeply in the broken phase at zero temperature. The
used quartic potential is motivated by the simple form of the classical
kink solution:
\begin{equation}
\Phi(x,y)=v\tanh\left(M x\right)\,
\end{equation}
with
\begin{equation}
M^2=-m^2/2\quad\textrm{and}\quad v=\sqrt{-6m^2/\lambda}\,.
\end{equation}
The tension of the wall is inversely proportional to the coupling:
$\sigma=4|m|^2/3\lambda$. 
In the classical limit the actual magnitude of the coupling is irrelevant as
it can be scaled out. In the numerics we used $\lambda=6M$. The only other
parameter, the mass sets the scale for the evolution, we use the inverse wall
width $M$ to render all variables dimensionless, this numerically means
$M=1$.

We discretise the model on a spatial lattice. We solve a cut-off theory
with a pre-set lattice spacing $a$. Since much of the physics of our
interest is in the infrared, $a$ plays little role. Based on earlier
numerical experience we can use lattices as coarse as $aM=0.5$.
We repeated the presented numerical analysis on a coarser lattice ($aM=0.7$)
and found no significant difference.
The lattice size $L$, however, matters. In order to avoid the interaction
of a pair of signals originating from the same site we stop the simulation
at $t=L/2$. This assumes that at $t=0$ there is no correlation between any of
the sites. Our initial condition will approximately satisfy this condition.

The initial condition can introduce other scales. We start the dynamics from a
low energy density random configuration with a rich domain wall structure. 
Our main interest is how these walls evaporate under the realistic
assumption that the scale in the initial condition separates from the
microscopic scale $M$.

We designed the following numerical experiment. We start from a
white noise configuration at $t=0$, deep in the symmetric phase. We also
checked the invariance of our results under replacing the initial noise by a
(classical) thermal equilibrium of the same energy density.
We then apply a cooling by adding a friction term to the kinetic term in the
equation of motion: $\partial^2\phi\to\partial^2\phi+2\gamma\partial_0\phi$.
This evolution is non-physical, and we switch off at a conveniently chosen time
when
the particle content is negligible and the domain wall density reached a
desirable value \footnote{We typically used a cooling of $\gamma=0.5$ in the
range $t=0\dots10$, followed by a relaxation period of 5 units. The
initial energy density before cooling was set to $2.3$.}.  We starting the
numerical observations only after a short period of relaxation after the
non-physical dynamics has been switched off. This corresponds to the
pre-thermalisation time scale \cite{prethermalization}.

The field configuration at this instant is the initial condition of the
dynamics of interest. We could set the origin of time to this instant, but
we choose not to. The $t=0$ point marks the onset of cooling, because at that point the
correlation length is known to be of the microscopic scale.


\subsection{Scaling solutions} 

The solution of classical dynamics is a straightforward computational task
and being restricted to two spatial dimensions our resources allows for
larger lattices ($L\ge1000$), too.

\begin{figure}[htp]
\centerline{\includegraphics[width=3in]{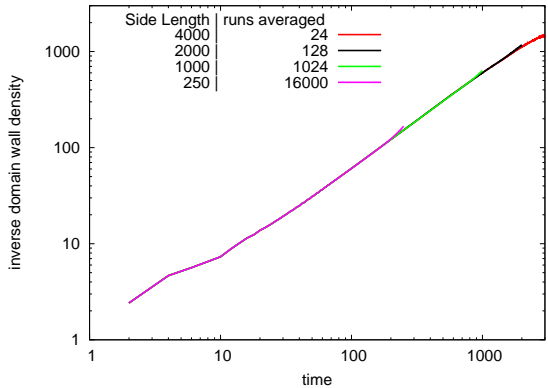}}
\caption{
Inverse domain wall density as a function of time. There is a perfect
linear correspondence over at more than two orders of magnitude
($t=10\dots2000$), from the time of ending the damped evolution at $t=10$ to
the end of the simulation, when the walls' mean curvature radius is over $10^3$
times larger than their width.
\label{fig:clwall}}
\end{figure}

As a first observation, we can estimate the domain wall length following
the techniques detailed in Ref.~\cite{Garagounis:2002kt}, and reproduced
their scaling solutions. In Fig.~\ref{fig:clwall} we show the inverse domain
wall density as a function of time. If the length scale ($\ell$) of the domain
wall network decouples from the microscopic scale, one may expect from
dimensional reasons: $\ell^{-1}\sim\L\sim t^{-1}$. The domain wall density
$\L$ we define as the total length of domain walls on the lattice divided
by the volume. A link on a lattice is considered as a part of a domain wall
if the sites at its both ends have field amplitudes of opposite sign.

As always in classical field theory we always display an average over an
ensemble of runs. (This is the thermal or white noise ensemble at $t=0$. At
positive times there is no randomness in the dynamics.) The averaged domain
wall densities start deviating near $t\approx L$, slightly later than expected.

This scaling is a manifestation of a more generic feature in classical
field theories, as it has been found in flat or curved space-time,
and in two or three dimensions \cite{Garagounis:2002kt}.

\begin{figure}[htp]
\centerline{
\includegraphics[width=1.5in]{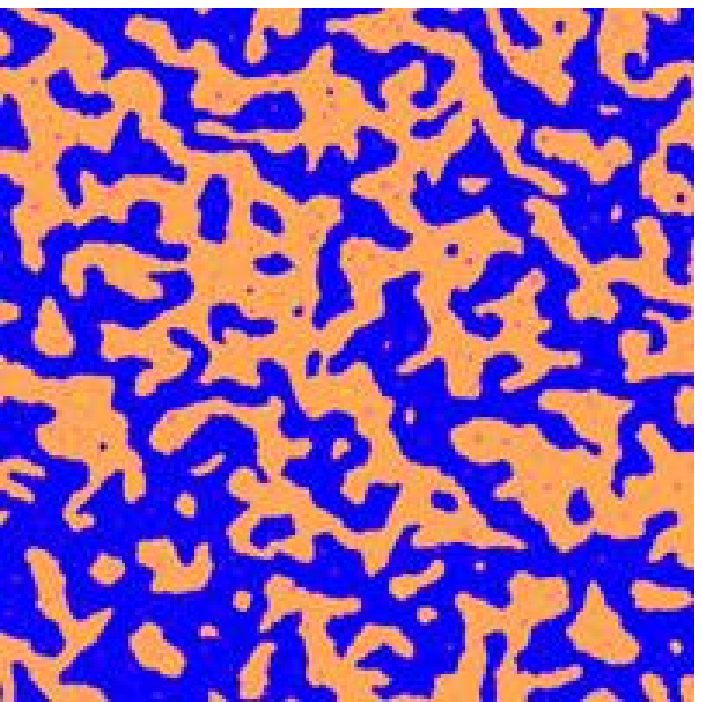}
\hspace{5mm}
\includegraphics[width=1.5in]{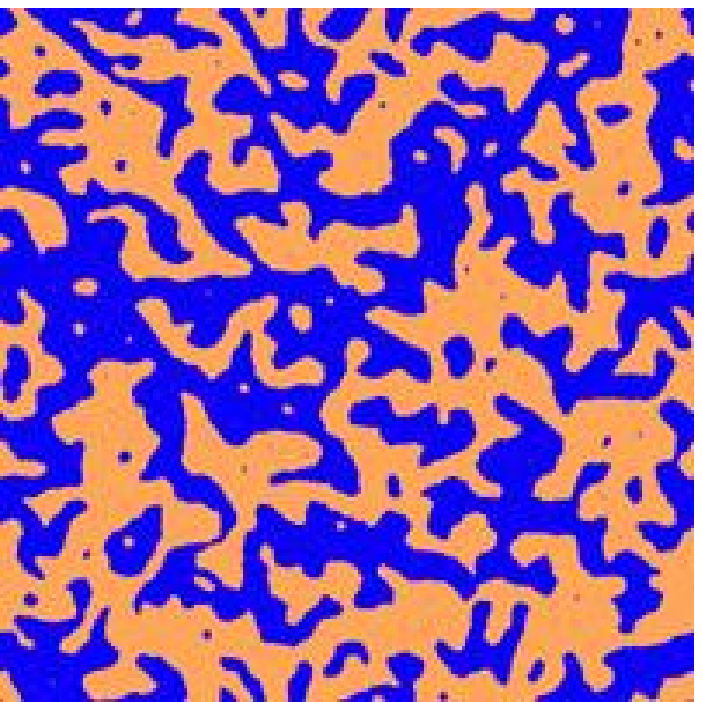}
}
\caption{
Two snapshots of the lattice field configurations ($L=2000$).
Blue and orange regions show the domains of the degenerate vacua.
To the left, we show the configuration at $t=50$, we cropped the region
$(0,L/2)\times(0,L/2)$ and scaled up by a factor of two. To the right
we show the uncropped lattice at $t=100$. There is no qualitative difference
between the snapshots. 
\label{fig:snapshots}}
\end{figure}

So that we gain more insight into the observed scaling we show a pair of
lattice configurations in Fig.~\ref{fig:snapshots}. The similarly looking
snapshots were taken at times 50 and 100, respectively, but the earlier
configuration we halved in linear size and scaled up accordingly.
The time evolution appears to be equivalent to zooming.

\begin{figure}[htp]
\centerline{\includegraphics[width=3in]{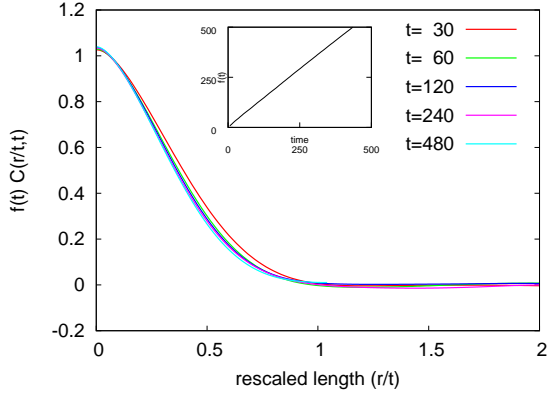}}
\caption{
Scaling of the equal time correlation function. Notice that the function
has an approximate Gaussian shape, which is due to the nature of spinodal
instability that created the domains. 
(There are deviations at small distance.) ($L=1000$, 1024
runs averaged)
\label{fig:cfn_scaling}}
\end{figure}

This feature can be made more formal in terms of the correlation function.
We define our $C(r,t)$ correlation function as
\begin{equation}
C(r,t)=\frac{1}{L^2}
\int dx dy dz \left\langle \phi(x,z,t)\phi(y,z+r,t)\right\rangle\,,
\end{equation}
where $\langle\cdot\rangle$ denotes an ensemble averaging. If the more
conventionally defined correlation function \footnote{
We prefer to use $C$ in the numerics for its simplicity and also because it
is the Fourier transform of the power spectrum. It simply relates
to $G$ if we assume isotropy and large volume. 
We only show data for times where these conditions are granted.
}
$G(|\vec x-\vec y|,t)=\langle\phi(\vec x,t)\phi(\vec y,t)\rangle$
scales as $G(r,t)=t^{\alpha} G(r/t)$, it is easy to see that
$C(r,t)=t^{\alpha+1} C(r/t)$.  Indeed, Fig.~\ref{fig:cfn_scaling} shows
a numerical evidence for the scaling of $C$, with $\alpha\approx 0$.
(If we require $t=0$ to be the origin, the scaling law is only accurate to
10 \%. If we fit the location of the origin of time and drop the initial
evolution ($t<200$), this we get an accuracy of 3\%. The displacement of
the origin fits to about $-26$ ).

\section{Hartree approximation of scalar fields} 
\label{sec:hartree}
In this section we review the inhomogeneous Hartree approximation and its
application to our model. The reader can find a more detailed introduction
in Ref.~\cite{Salle:2000hd}. Its other name, Gaussian approximation,
reflects the essence of the truncation of the dynamics: we disregard any
connected higher $n$-point functions.  Note in the context of the $N$-component
scalar field the leading order in $1/N$ expansion leads to very similar (also
Gaussian), but inequivalent approximation \cite{Mihaila:2000ib}.


The operator equation in Heisenberg picture that we have to solve reads:
\begin{equation}
(\partial^2+m^2)\hat\phi\xt+ \frac{\lambda}{6}\hat\phi^3\xt=0\,.
\label{eq:Heisenbergeom}
\end{equation}
We split off the quantum expectation value:
$\hat\phi\xt=\bar\Phi\xt+\phiop\xt$, with $\avr{\phiop\xt}\equiv0$.

The Hartree approximation basically means that we restrict the density
operator of the $\phiop$ degrees of freedom to Gaussian at all times.
For such degrees of freedom $\phiop_i$ we have the following identities:
$\avr{\phiop_i^3}=0$ and $\avr{\phiop_i^3\phiop_j}=
3\avr{\phiop_i^2}\avr{\phiop_i\phiop_j}$.

We can simply take the quantum average of Eq.~(\ref{eq:Heisenbergeom}),
or multiply from the left by $\phiop(\y,t_y)$ and get an equation for
the Wightman propagator $G^<(\x,t_x;\y,t_y)\equiv G^<(x,y)$ by averaging that, too:
\begin{eqnarray}
\left[\partial^2_x+m^2 + \frac{\lambda}{2}G^<(x,x)\right]
\bar\Phi(x)+\frac{\lambda}{6}\bar\Phi^3(x)&=&0
\label{eq:H1}
\,,\\
\left[\partial^2_x+m^2 + \frac{\lambda}{2}\bar\Phi^2(x)
+\frac{\lambda}{2}G^<(x,x)\right]
G^<(x,y)&=&0
\,,
\label{eq:H2}
\end{eqnarray}
It is remarkable that this simple truncation of the hierarchy of $n$-point
functions leads to a self-consistent set of equations. Indeed, these are
the Schwinger-Dyson equations for the propagator in the leading order
truncation of the two-particle irreducible (2PI) effective action
\cite{Cornwall:1974vz}.


Gaussianity also implies that the Heisenberg operators at a finite time relate
to the initial operators by a Bogolyubov transformation: \begin{equation}
\phiop\xt=\intk\left(\ak\psik\xt+\ak^+\psik^*\xt\right)\,.  \end{equation} So
that we have the text-book operator at $t=0$ we set
$\psik(\x,0)=e^{-i\kx}/\sqrt{2\omega_k}$ and
$\dot\psik(\x,0)=-i\omega_k\psik(\x,0)$. Here $d$ is the number of space
dimensions and $\omega_k^2=\vec k^2+m_r^2$, where $m_r$ is the renormalized mass,
often amended with a contribution from the background field. The ladder
operators obey the usual
$[\hat a_{\vec k_1}^{\,},{{\hat a}_{\vec k_2}}^+]
=(2\pi)^d\delta(\vec k_1-\vec k_2)$
commutation relation. The initial particle spectrum are given by
$\avr{\ak^+\ak}=n^0_{\vec k}$. These numbers appear in the equal time
two-point function:
\begin{equation}
G^<(\vec x,t,\vec x,t)=\intk |\psik\xt|^2(2n^0_{\vec k}+1)
\label{eq:tadpole}
\end{equation}
This obviously diverges even in two space dimensions. The initial infinite
mass shift we compensate by a mass renormalisation and introduce the finite
mass squared $m_r^2$ in the equation for $\psik$:
\begin{eqnarray}
\Bigl[\partial_x^2+m_r^2+\frac{\lambda}{2}\bar\Phi\xt&&\nonumber\\
+\frac{\lambda}{2}
  \intk\left[|\psik\xt|^2-\psik(\vec x,0)|^2\right](2n^0_{\vec k}+1)&&
\label{eq:MFE}\\
\Bigr] \psi_{\vec p}\xt&=&0\nonumber
\end{eqnarray}

A coupling renormalisation is also necessary in three dimensions
\cite{BaackeRenorm}, but in this simple case we do not need to go beyond
mass renormalisation.

The traditional way of solving the dynamics in Gaussian approximation
involves Eqs.~(\ref{eq:H1}), (\ref{eq:tadpole}) and (\ref{eq:MFE}).
One normally discretizes the equations on a space lattice. A consistent
Bogolyubov transformation requires that the $\vec k$ index of the mode functions
runs in the entire Fourier space of the lattice. In addition to the trivial
background equation on an $N^2$ lattice this means $N^4$ complex equations.


But mode function expansion is just one of the possible ways of solving
Eqs.~(\ref{eq:H1}) and (\ref{eq:H2}). Alternatively, we consider an ensemble
of $N_e$ classical trajectories $\varphi_i\xt$, solutions of the equation
\begin{equation}
\left(
\partial^2_x+m^2+
\frac{\lambda}{2}\left[\bar\Phi^2\xt+\avr{\varphi^2\xt}_E\right]
\right)\varphi_i\xt=0\,.
\label{eq:linphi}
\end{equation}
Here $\avr{\cdot}_E$ stands for the ensemble average. Indeed, multiplying
the equation with $\varphi_i(\vec y,t_y)$ and averaging over $i$
(ensemble average), will bring us back to Eq.~(\ref{eq:H2}) with
$G^<\to G^e=\avr{\varphi(\vec x,t_x)\varphi(\vec y,t_y)}_E$. But there
will be no exact equivalence between ensemble and quantum averages: the
quantum two-point function $G^<$ is complex, $G^e$ is real. Notice, however,
that the imaginary part of $G^<$ entirely decouples in Eq.~(\ref{eq:H2}),
since the equal time propagator is always real.

Of course, $\varphi_i\xt$ must be properly initialized to form a
Gaussian ensemble of the correct standard deviation:
\begin{eqnarray}
\avr{\varphi(\vec x,0)\varphi(\vec y,0)}&=&
\hbar
\intk e^{-i\vec k(\vec x-\vec y)}
	\frac1{\omega_k}\left(n^0_{\vec k}+\frac12\right)
\,,\nonumber\\
\avr{\dot\varphi(\vec x,0)\dot\varphi(\vec y,0)}&=&
\hbar
\intk e^{-i\vec k(\vec x-\vec y)}
	\omega_k
	\left(n^0_{\vec k}+\frac12\right)
\,,\nonumber\\
\avr{\varphi(\vec x,0)\dot\varphi(\vec y,0)}&=&0\,.
\label{eq:fluctic}
\end{eqnarray}
Technically, we initialize $\varphi$ in momentum space by
a random phase and amplitude at the $t=0$ and $t=\delta t$
time slices.

We intentionally introduced the factor $\hbar$, as a control
parameter for the fluctuation $\varphi$. This way we can
tune strength of the back reaction of the quantum fluctuations
to the background. In the classical theory it was possible to
scale out $\lambda$, here rescaling the field with $1/\sqrt{\lambda}$
would also require to rescale $\hbar$ with $\lambda$. If we stick
to $\lambda=6M$ in the numerics, it is the $\hbar$ in Eq.~(\ref{eq:fluctic})
that one can use to vary the coupling, effectively.

Numerically, it is much simpler to solve $N_eN^2$ real equations than
$N^4$ complex ones, we found that even an ensemble of $1\ll N_e\ll N^2$ was big
enough. The simple structure of Eq.~(\ref{eq:linphi}) allows high speed
implementations \footnote{
Two of the important optimizations: we use the single precision $SSE$
extension of the i386 architecture and that we loaded $\varphi_i\xt$
into the cache once in a time step, only.}.
We note that the equation is not stable without manually fixing
$\avr{\varphi_i(\vec x)}=0$ after every leap-frog time step.


Before embarking into the analysis of numerical results, let us pause to 
discuss in what sense the Hartree equations represent a quantum
correction to the classical dynamics.

Notice that we can arrive at Eqs.~(\ref{eq:H1}) and (\ref{eq:H2}) also from
a different concept. Let us start a number of classical
trajectories from an initial Gaussian ensemble (e.g. Eq.~(\ref{eq:fluctic})).
Instead of following the individual trajectories we can write down the
equations for the $n$-point functions. Simply discarding the three or
higher order correlators we get a closed set of equations, that coincide
with Eqs.~(\ref{eq:H1}) and (\ref{eq:H2}). We would also arrive
to the same equations by truncating the 2PI effective action for the classical
(or quantum) field theory to leading order.

Indeed, whether we start from a classical or quantum Gaussian ensemble,
the genuine quantum features start to appear if we keep the four-point
equation at least. A self-consistent set of equations follows from the
next-to-leading order truncation of the 2PI effective action, where
one easily identifies the term, responsible for quantum effects
\cite{Aarts:2001yn}.

This statement, however, means that to Hartree order it is only the initial
condition that reflects quantum physics. Do the mode function equations
(\ref{eq:MFE}) or the propagator equation (\ref{eq:H2})
introduce quantum corrections at all?

The answer is yes. If we consider one single classical trajectory $\bar\Phi$,
switching on $\hbar$ in Eqs.~(\ref{eq:fluctic}) will definitely enable
many quantum phenomena, such as vacuum particle production. Instead of
doing Hartree, one can, of course, consider an ensemble of $\bar\Phi$ fields,
initialized (as usual) with the just-the-half rule (analogous to
Eqs.~\ref{eq:fluctic}) and evolve them classically.  This classical ensemble
will equally enable the same quantum phenomena, but it will bring in several
classical artefacts, too, such as the decay of the quantum zero-point energy.
These artefacts can most simply eliminated by shutting down all higher loop
diagrams, down to the order where quantum and classical approximations agree:
this is the Hartree approximation.

Although it is possible to properly include higher order corrections
\cite{Berges:2002cz,Berges:2001fi}, they are not inevitable in the following
two extremes: If the particle numbers are low, the higher order quantum
corrections, that account for scattering of the quantum fluctuations, are not
very important compared to the dynamics of other energetic objects, such as
defects. If the particle numbers are very high, higher order quantum
corrections are crucial, but they can be well estimated by a
classical ensemble, here $n_{\vec k}$ dominates in $n_{\vec k}+1/2$ and the
classical artefacts will then be suppressed.

In the application considered in this paper we work with low particle numbers
produced by the sparse network of defects. We believe that we do not miss
the magnitude of the particle's back-reaction by ignoring the thermalisation of
their spectral distribution. This semiclassical approach, however,
approximates the damping of the classical degree of freedom at the
lowest order.

\section{Classical versus Hartree dynamics} 
\label{sec:comparison}
The initial conditions given in section \ref{sec:model} define a (highly
non-Gaussian) ensemble of domain wall configurations at $t=15$. For each member
$\bar\Phi_i$ we define a (Gaussian) sub-ensemble of fluctuations. We follow the
dynamics of this sub-ensemble in the Hartree approximation. The final averaging
over the domain wall configurations occurs at the very end of the calculation.
At the time we switch on the quantum equations (\ref{eq:linphi}) we renormalize
the mass and thereby allow a smooth transition to quantum evolution.

\begin{figure}[htp]
\centerline{\includegraphics[width=3in]{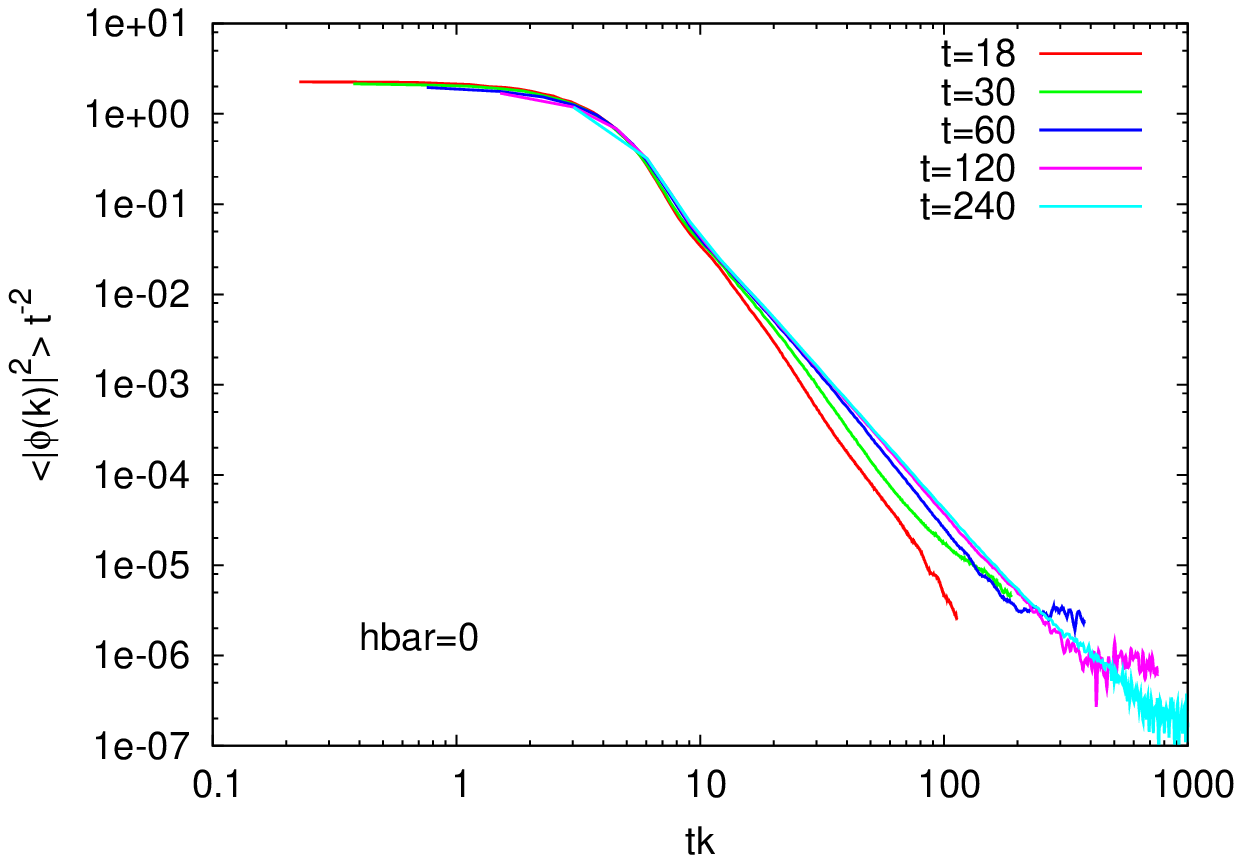}}
\centerline{\includegraphics[width=3in]{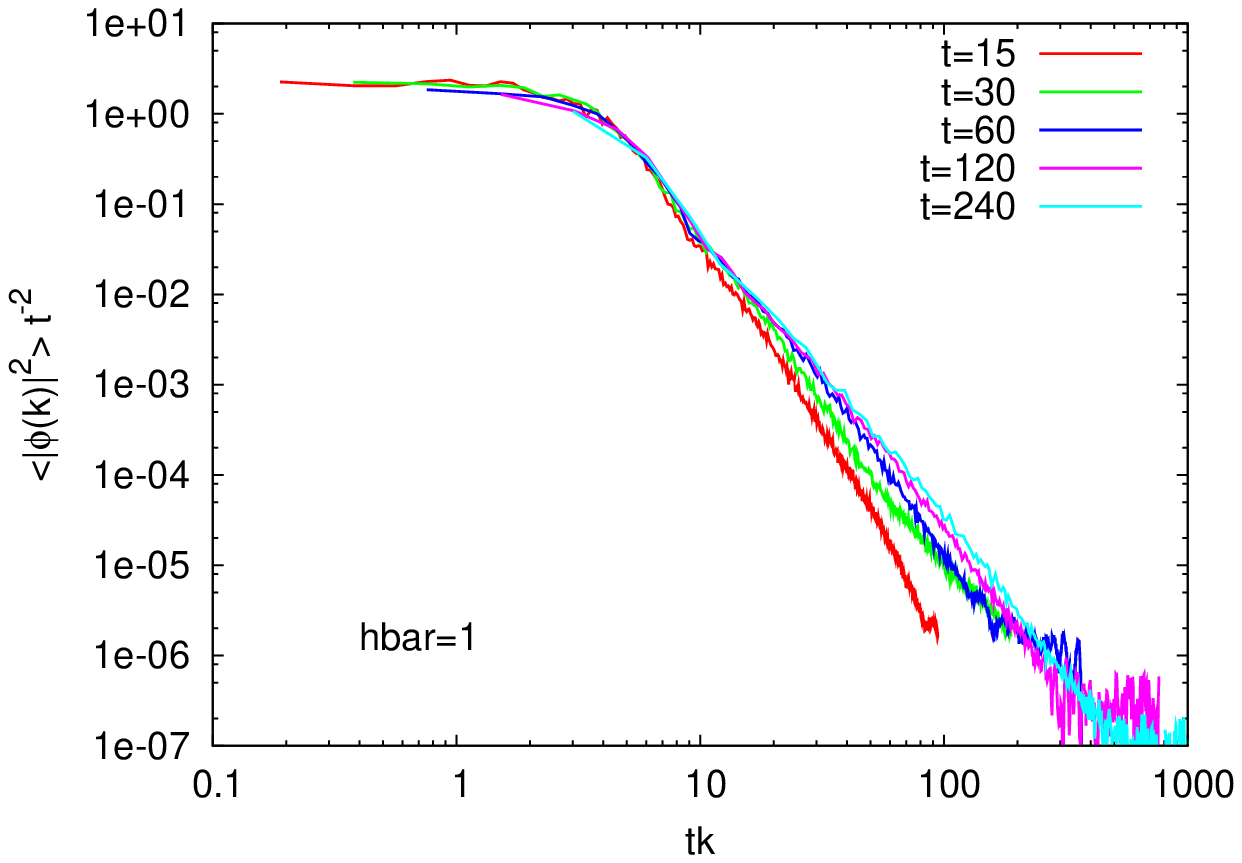}}
\caption{
The scaling of the power spectrum. The macroscopic part (domain walls) clearly
follows the scaling law both in the classical (top) and in the quantum (bottom)
case.  The scaling breaks at the tail of the spectrum
(small classical structures). ($L=500$, average of 16000 and 192 runs
for $\hbar=0$ and $1$, respectively.) 
\label{fig:ps}}
\end{figure}

In Fig.~\ref{fig:ps} we show the evolution of the power spectrum of the
background field. In classical field theory this is the only degree
of freedom, whereas in the Hartree approximation energy may drift into
the ``modes'' (the ensemble of quantum fluctuations). 

\begin{figure}[htp]
\centerline{\includegraphics[height=2in]{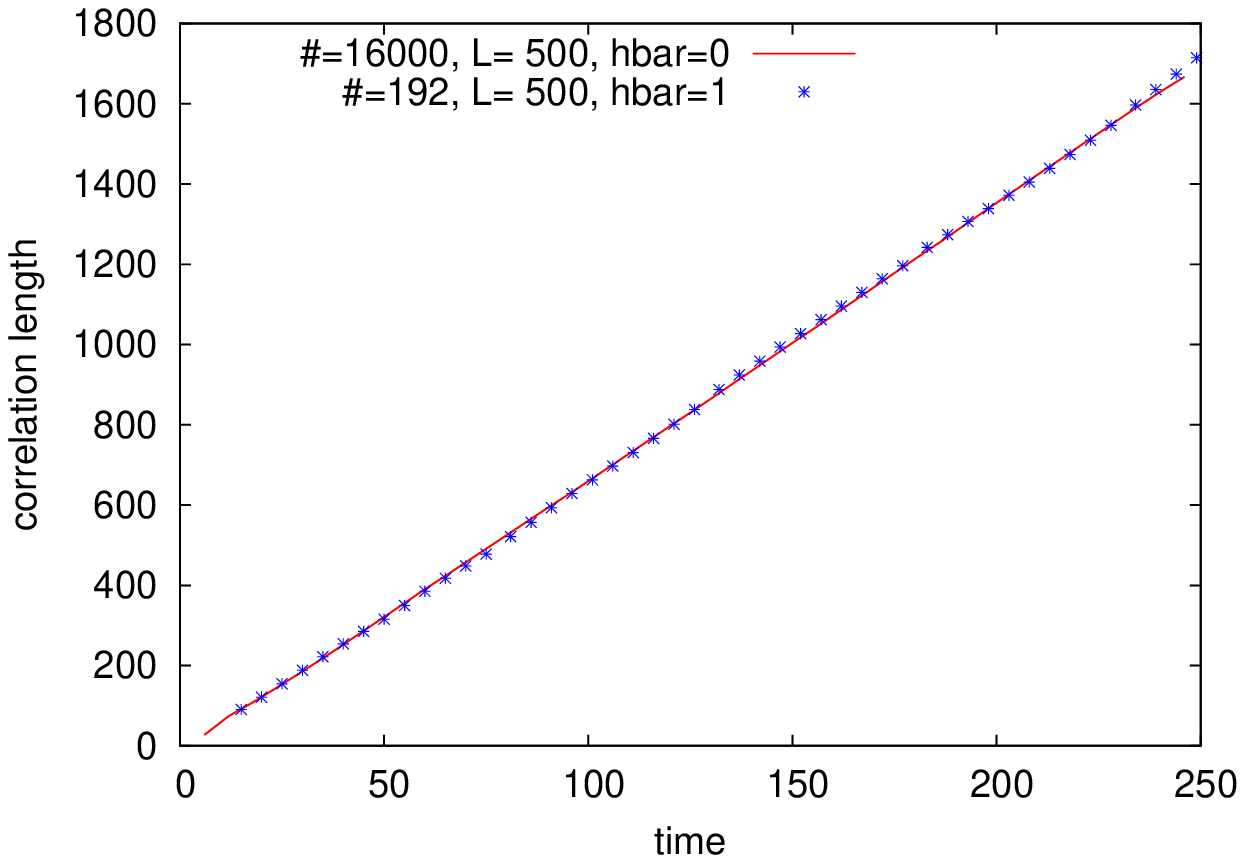}}
\centerline{\includegraphics[height=2in]{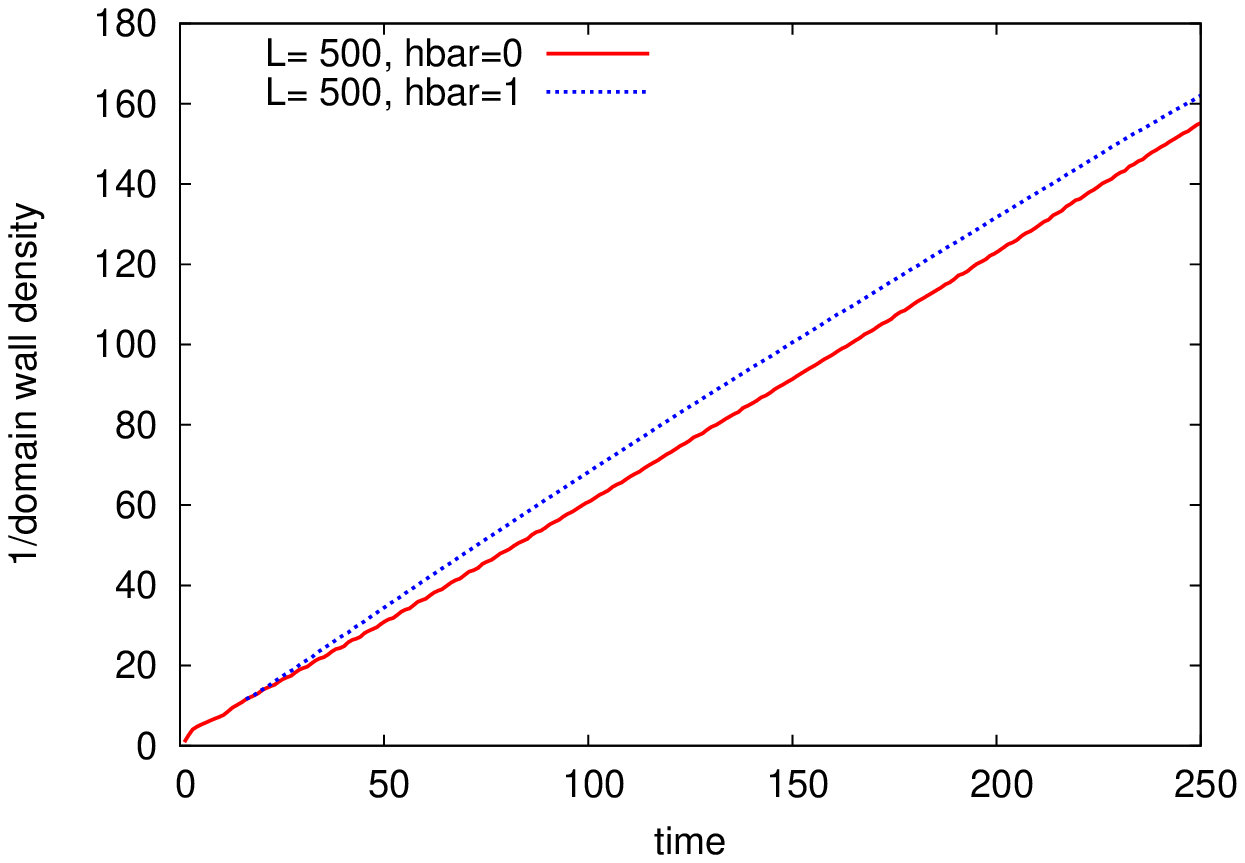}}
\caption{
The scaling of the correlation length (top) and the domain wall density (bottom)
\label{fig:corr}}
\end{figure}

The correlation length in Fig.~\ref{fig:corr} is defined by a Gaussian
fit to the correlation function shown in Fig.~\ref{fig:cfn_scaling}.
In harmony with Fig.~\ref{fig:ps} we see no impact of the quantum fluctuations
on the evolution of the macroscopic degrees of freedom. If all the domain
wall loops were macroscopic, this would suggest that inverse total length of
domain walls shown in Fig.~\ref{fig:corr} receives no significant quantum
correction.  Indeed, we again find a linear scaling, and we could not
find a significant correction to the slope parameter for $t>150$. But
the inverse domain wall density is not entirely linear. In the first half
of the evolution it drifts away the classical solution. This reflects
a transient decay of some classically more stable structures.

The power spectrum in Fig.~\ref{fig:ps} does not give account on the
created particles. The power spectrum of the $\varphi_x\xt$ functions
in Eq.~\ref{eq:linphi} reflect the created particles. This spectrum does
not scale, and performs a ``boring'' evolution: only the amplitude changes
slightly and always resembles the vacuum power spectrum. This confirms the
assumption that the particles are created on the mass scale and not e.g.
in the infrared. At and beyond the mass scale the quantum fluctuations
dominate over the classical structures in the power spectrum.

By construction of the initial conditions the energy density transferred from
the domain walls to the fluctuations does not significantly raise the
temperature and so the thermal mass. It was an important assumption in our
analysis that the wall width $M$ is constant in time.

The domain wall density and the correlation length are the key observables
when we discuss scaling. Irrespectively to the coupling ($\lambda$) or
the strength of the fluctuations ($\hbar$) we fit $\xi\approx 3.4(2)\cdot t$.
For the domain wall density we find ${\cal L}=l/L^2\approx 1.66(3)/t$ 
for our initial condition, where $l$ is the total counted length of domain
walls on the lattice at a given time. It is remarkable that these dimensionless
coefficients are robustly insensitive to the variation of the coupling or
the value of $\hbar$. Also it does not depend on the lattice spacing nor
on the initial noise amplitude or the details of the cooling procedure.

\begin{figure}[htp]
\centerline{\includegraphics[width=3in]{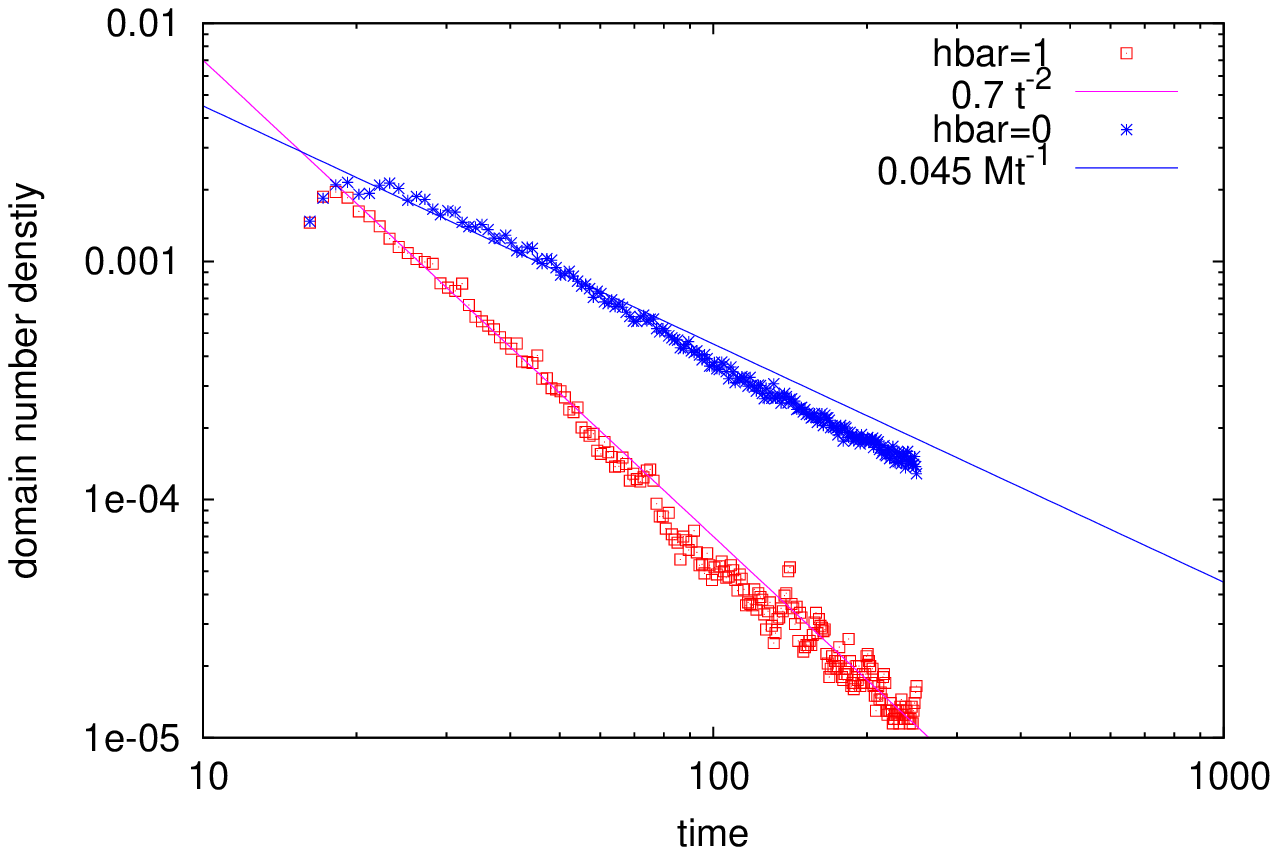}}
\centerline{\includegraphics[width=3in]{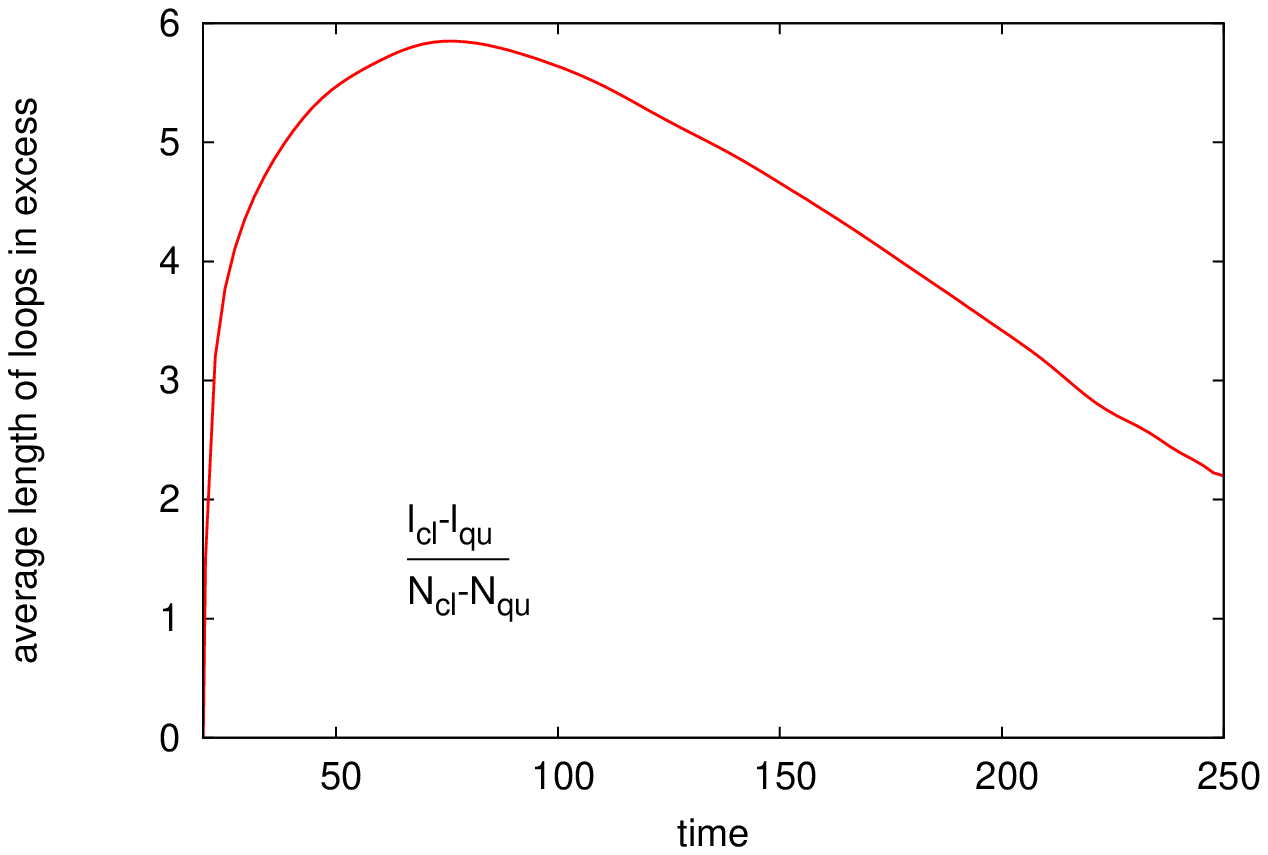}}
\caption{
Number of domains over lattice volume (top) in the classical and quantum
framework. The classical scaling is counter intuitive. Dividing the
difference of the total loop length by the number of loops (estimated by
the number of domains) we get an average loop size (bottom).
\label{fig:number}}
\end{figure}

To gain more insight into the small discrepancy between the quantum and
classical domain wall density we count the number of domains, and use this
number to estimate the loop number density ($n(t)=N(t)/L^2$). We applied a
cluster algorithm on the lattice and plotted the resulting number density in
Fig.~\ref{fig:number}. The effect of the quantum fluctuations is now striking.

In the approximated quantum theory we get what we think we should: if
the correlation length $\xi\sim t$ scales linearly, any number density
must scale as $n_{q}\sim t^{-2}$. The conclusion from Fig.~\ref{fig:number} is
that in the classical theory the domain number is dominated by microscopic
structures. One hint for the smallness of these ``mini-domains'' is that the
inverse wall width $M$ must appear in the $n_{cl}(t)\sim M/t$ scaling rule
for dimensional reasons, and its coefficient is not extreme. We can directly
measure an average defect loop size by counting the loops (or domains,
practically) that are in excess in the classical solution. The total wall
length is also bigger in the classical case than with quantum correction.  We
used their quotient to estimate the size of these loops in
Fig.~\ref{fig:number}. Unfortunately our numerics is not conclusive at later
times, we expect this ratio to settle at a positive value $\sim M^{-1}$.

\begin{figure*}[htp]
\centerline{
\includegraphics[width=7in]{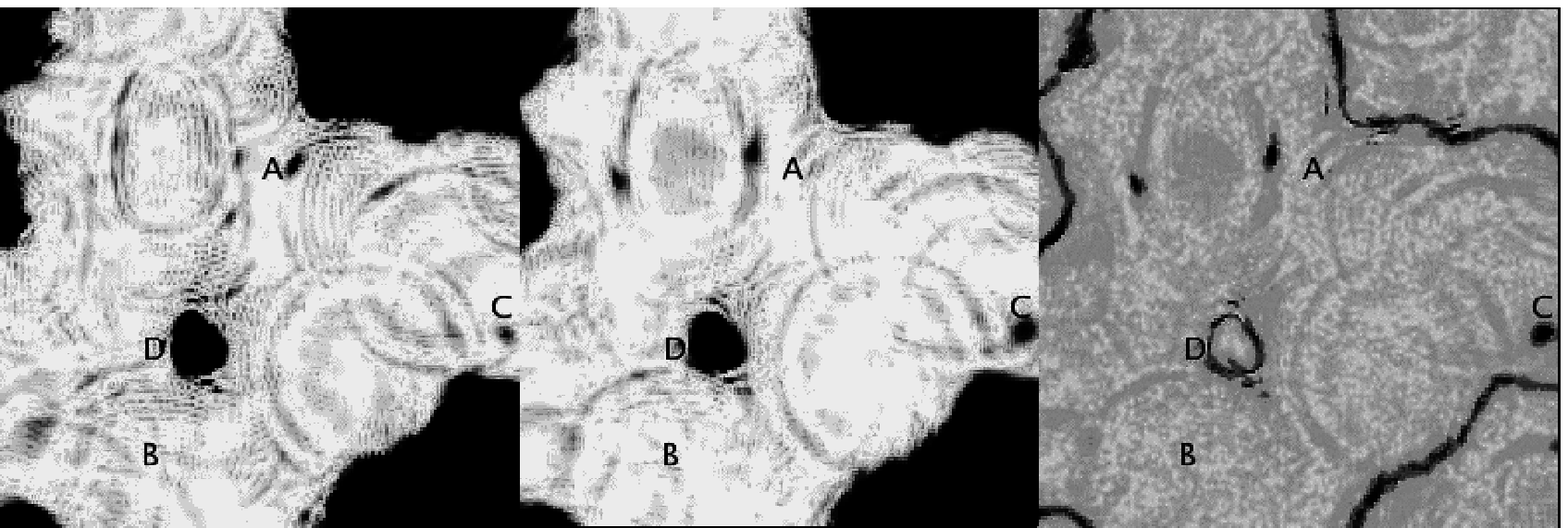}
}
\caption{
Snapshot from a classical (left) and the corresponding quantum (middle)
run. To the right the particle content is characterized by
$\avr{\varphi^2\xt}_E$.  The darker points mean higher value. (These images
were taken at $t=40$ on a $L=128$ lattice. We cropped out a piece of
$75\times75$)
\label{fig:cmpsnapshot}}
\end{figure*}

To find out more about these small classical structures let us compare
the lattice snapshots taken from the same run with and without quantum
fluctuations. We picked the time $t=40$ and cropped a larger lattice
appropriately so that we can show the most phenomena in one image:
Fig.~\ref{fig:cmpsnapshot}

In these images the black regions correspond to the $\bar\Phi>0$ domains. We
plotted the $\bar\Phi<0$ domain with better colour resolution. The left
hand side plot corresponds to the classical evolution, it shows many
ripples and numerous dark spots where $\bar\Phi$ goes locally close or beyond
zero. In the middle (quantum) plot we see fewer ripples and some (but not all)
of the dark spots of the classical plot are missing here. To the right, we
demonstrate the excitation of the quantum modes by plotting
$\avr{\varphi^2\xt}_E$.
These fluctuations are the strongest on the domain walls, which we interpret 
as particles in the bound state.

Looking at a sequence of such snapshots gives more detail about how these
ripples are produced. As the domain walls shrink they emit classical waves,
with a wave length of few times the wall width. These waves are equally
present in the quantum case as well, where they are more damped. In the
quantum fluctuation plot we also find traces of the classical ripples, but
the spatial distribution of the produced particles appears to be smooth,
and the ripples in $\avr{\varphi^2\xt}_E$ are an order of magnitude smaller
than in $\bar\Phi^2$.

On the snapshots we marked the interesting places by letters. In
both sides of the letter ``A'' the ripples are locally so high in amplitude
that these spots are counted as a domain by the cluster algorithm
and they contribute to the total length of domain walls. But they are not
counted into the statistics of walls in the quantum case, since then their
amplitude is within the threshold of zero field value. The amplitude of
these spots actually oscillates, this is why we do not see the one on the
right hand side of the letter ``A'' in the quantum plot.
``B'' marks the centre of ripples emitted earlier by the collapsing bubble
marked with ``D''. These are mostly damped in the quantum run. The waves
around the bubble ``D'' are higher in amplitude than in the quantum case.
Finally, there is a spot with strongly oscillating amplitude, marked with
``C''.  The magnitude of the quantum fluctuations oscillate coherently with the
background field value.

\section{Discussion}
\label{sec:discussion}
Let us summarize the numerical findings: The correlation length, which
is fitted from the correlation function in direct space, reflects the
macroscopic evolution. We find that the known scaling behaviour is
unperturbed by quantum effects. On the microscopic level, however,
where the scaling is broken, we find stronger quantum effects, as expected.

We also find, that there are ``mini-domains'' in the classical simulation,
that (at least partly) disappear if we switch on the quantum degrees of
freedom. Its simplest explanation is that there are classically stable small
structures that decay in a quantum field theory. Now we can speculate
what these could be. Natural candidates are oscillons, localised
oscillating wave packets, which are (quasi-)stable solutions of the classical
field theory \cite{Copeland:1995fq,Hindmarsh:2006ur}. 

If these small structures are indeed oscillons, their stability is enhanced
by low dimensionality. If in three dimensions
they are subject to a more rapid decay \cite{Fodor:2006zs,Saffin:2006yk}, 
making the quantum decay
channel less significant and hence the quantum correction to the
scaling even smaller.

Indeed, a closer look on the lattice field revealed that there are
small regions (with a diameter of O(5) domain wall width) that oscillate
with a frequency $\sim M$. But oscillons are not the only structures that
appear. The shrinking and collapsing domain walls emit classical waves
with a wave length $\sim M^{-1}$. We see these waves on the
lattice snapshots as circular ripples. These ripples from various sources
interfere and at the points of constructive interference the field value
may locally exceed zero and will then be counted as a small domain.

Classical waves are emitted in the quantum field theory, too.
In quantum mechanics this classical excitation is known as coherent state,
which transforms into an enlarged width of the wave function, or
into particles in field theory language. This is the
point where quantum corrections enter:  the classical waves are damped
and their interference results in fewer and less stable localised oscillating
wave packets.

In this picture there is a non-perturbative classical
mechanism that converts the energy stored in the string (or domain wall)
to microscopic objects. In a field theory, these objects are neither
loop fragments, nor particles, but coherent oscillations of the field
expectation value. Our numerics suggests that the scale of these classical
waves are on the microscopic scale $M$. We observe that these waves are 
emitted from structures of size $\ell$, presenting us with the challenge of 
explaining energy transport over a huge scale separation,  
$M\ell \sim 10^3$ at the end the simulation.

It is clear from the shown numerics that the domain wall decay was not
enhanced by the quantum fluctuations and this conclusion we checked
to stay true with $\hbar=2$ or $\lambda=12$. There is no indication for
a direct decay channel into particles. 
A direct decay might also manifest in the sensitivity to the choice
of the lattice spacing as we switched between $aM=0.5$ and $0.7$, but
we found no significant difference. However, the decay of 
the classical waves and oscillons is no
longer protected by scale separation. 

Finally, let us attempt to understand Fig.~\ref{fig:number}. The energy
density associated to macroscopic $D$-dimensional defects in $d$ dimensions
is $\sim M^{1+D}t^{D-d}$. Their decay
releases energy at a rate of $\sim M^{1+D}t^{D-d-1}$. This energy is used to
produce high amplitude classical structures (e.g. oscillons) that may have
been counted as small domains. Since they emerge on the microscopic scale,
their number density has a source of $C_{source}M^{D} t^{D-d-1}$, 
where $C_{source}$ and the other constants we introduce here are dimensionless
numbers of ${\cal O}(1)$.
These small structures can decay in various ways: a) In the quantum calculation
we include the direct quantum mechanical decay into particles with a rate of
$\Gamma\sim M$; b) The small objects can be hit by a domain wall or string, its
rate is proportional to the defect density: $C_{defect}M^{D-d+1}t^{D-d}$; 
c) These objects can also hit each other and
annihilate. The probability of a given small object to meet an other one is
proportional to its density $n$, which gives a rate of $C_{coll}M^{1-d}\,n$.
These together give the following equation for the density $n$
\begin{equation}
\dot n + \Gamma n + C_{defect}\frac{M^{D-d+1}}{t^{d-D}} n 
+ C_{coll}M^{1-d}n^2=C_{source}\frac{M^D}{t^{d-D+1}}
\label{eq:nsmall}
\end{equation}
If the quantum decay into particles dominates giving a finite life time to
these small classical structures, the density $n$ simply follows the source.
Indeed we see $n\sim t^{-2}$ in Fig.~\ref{fig:number}. In the absence of
$\Gamma$, however, we find that $n\sim M/t$ solves Eq.~(\ref{eq:nsmall})
in consistence with our observation. Since in this case $n$ shows the same
scaling as the domain wall density, counting them as defects does not spoil the
observation of scaling.  The classical approximation of Eq.~(\ref{eq:nsmall})
suggests that for $d>2$ the collision term would dominate, giving $n\sim
t^{-2}$.  In higher dimensions, however, oscillons and other analogous
structures are less stable, which introduces a decay term of classical nature
bluring difference between classical and quantum scaling.
\section{Conclusion}
\label{sec:conclusion}
In this paper we integrated the classical field equations as well as
the Hartree approximated quantum evolution of a
scalar field in the broken phase, starting from a network of domain walls.
The scaling of macroscopic observables was manifest also in the quantum
theory. Our numerical results suggest that the direct decay of domain
walls into particles is insignificant, as the perturbative estimates predict.
We can instead attribute the decay to the emergence of classical waves and other
structures, such as oscillons.  Since these coherent excitations
of the quantum field theory are produced at the microscopic scale, their
perturbative decay is no longer suppressed by the separation of scales.
The production of large-amplitude classical oscillations is a genuine 
non-perturbative phenomenon that deserves further investigation, as a 
similar effect is seen to drive  
the decay of cosmic strings in three dimensional field theory simulations 
\cite{Vincent:1997cx}.
Understanding the dominant decay channel of strings is of crucial 
importance for computing their observational signals.

\begin{acknowledgments}
The authors thank J\"urgen Berges and Anders Tranberg for the encouraging
discussions and also acknowledge the collaboration with Petja Salmi on a
related project. The numerical work has been carried out on the Archimedes
cluster of the University of Sussex. SB is funded by STFC.
\end{acknowledgments}

\end{document}